\documentclass[11pt]{article}%
\usepackage[latin1]{inputenc}%
\usepackage[english]{babel}
\usepackage{epsfig}%
\usepackage{amsmath}%
\usepackage{amsfonts}%
\usepackage{mathrsfs}%
\usepackage{amssymb}%
\usepackage{graphicx}%
\usepackage{longtable}%
\usepackage{geometry, calc, color, setspace}%
\usepackage{indentfirst}%
\usepackage{wrapfig}%
\usepackage{natbib}
\usepackage{lscape}
\usepackage{subcaption}
\usepackage{footnote}

\begin{document}

\begin{center}
{\Large{\bf Modeling cure fraction with frailty term in latent risk: \\ a Bayesian approach}}
\end{center}

\vspace*{0.5cm}

\begin{center}
Agatha Sacramento Rodrigues$^{1,2}$ \\
Vinicius Fernando Calsavara$^{3}$ \\
Vera Lúcia Damasceno Tomazella$^{4}$ \\
\end{center}

\vspace{0.5cm}
\noindent $^1$ University of S\~ao Paulo, Institute of Mathematics and Statistics, S\~ao Paulo, SP, Brazil.\\
$^2$ S\~ao Paulo University Medical School, Department of Obstetrics and Gynecology, S\~ao Paulo, SP, Brazil. \\
$^3$ A.C.Camargo Cancer Center, Department of Epidemiology and Statistics, International Research Center, S\~ao Paulo, SP, Brazil. \\ 
$^4$ Federal University of S\~ao Carlos, Department of Statistics, S\~ao Carlos, SP, Brazil.\\
$\ast$ E-mail:  agatha.srodrigues@gmail.com; vinicius.calsavara@cipe.accamargo.org.br; vera@ufscar.br

\vspace*{1.5cm}

\thispagestyle{empty}

\noindent \textbf{Abstract}: In this paper we propose a flexible cure rate model with frailty term in latent risk, which is obtained by incorporating a frailty term in risk function of latent competing causes. The number of competing causes of the event of interest follows negative binomial distribution and the frailty variable follows power variance function distribution, in which includes other frailty models such as gamma, positive stable and inverse Gaussian frailty models as special cases. The proposed model takes into account the presence of covariates and right-censored survival data suitable for populations with a cure rate. Besides, it allows to quantify the degree of unobserved heterogeneity induced by unobservable risk factors, in which is important to explain the survival time. Once the posterior distribution has not close form, Markov chain Monte Carlo simulations are consider for estimation procedure. We performed several simulation studies and the practical relevance of the proposed model is demonstrated in a real data set.

\vspace*{0.5cm}


\newpage

\section{Introduction}

In survival data, an usual interest is to model the time until the occurrence of a defined event. In the traditional approach, it is assumed that all units under study are susceptible to the event of interest that will occur. However, such assumption can be violated because many studies have what we call ``immune'' or ``cured'' elements. The idea is that the event will never occur for immune units because they are not susceptible to the event of interest. Thus, a class of models referred as the cure rate models considers that situation and it has been studied by several authors. The Berkson-Gage model \cite{Berkson} was probably the first model to propose the cured fraction. This model is based on the assumption that only one cause is responsible for the occurrence of an event of interest \cite{cooner2007flexible}.

In biomedical studies, an event of interest can be the patient's death as well as cancer recurrence, which can be attributed to different latent competing causes as the presence of an unknown number of cancer cells. These causes are based on the fact that each surviving carcinogenic cell can be characterized by an unknown time (promotion time) during which the cell could become a definitive tumor \cite{cobre2013mechanistic}. The literature on this subject is extensive. The books of \cite{maller1996survival, Ibrahim}, as well as the articles \cite{Yakovlev, Chen, tsodikov2003estimating, yin2005cure, cooner2007flexible, Rodrigues, rodrigues2011destructive, rodrigues2012bayesian, rodrigues2015latent, castro2009bayesian, cancho2011flexible, cancho2012bayesian, cancho2013destructive, cancho2013long, borges2012correlated} could be mentioned as key references.

In the competing causes scenarios, the promotion times are usually assumed to be independent and identically distributed, i.e., the carcinogenic cells lifetimes follow a common distribution function and the most common choices have been exponential, piecewise exponential, Weibull, among other. Besides, the cure rate models implicitly assume a homogeneous population for the susceptible units. However, covariates can be included in the model in order to explain some heterogeneity. But there is an unobserved heterogeneity induced by unobservable risk factors, which are not considered in the model.

The models that take into account the unobservable heterogeneity are known as frailty models \cite{Vaupel_79}. These models are characterized by the inclusion of a random effect, that is, an unobservable random variable that represents the information that can not be observed, such as unobservable risk factors. If an important covariate was not included in the model, this will increase the unobservable heterogeneity, affecting the inferences about the parameters in the model. This way, the inclusion of a frailty term can help to relieve this problem.

The frailty term can be included in an additive form in the model. However, a multiplicative effect on the baseline hazard function is often used. Multiplicative frailty models represent a generalization of the proportional hazards model introduced by \cite{Cox_72}, which the frailty term acts multiplicatively on the baseline hazard function. This approach has been studied by several authors, notably \cite{Clayton_78, Vaupel_79, andersen_1993, hougaard_1995, sinha_1997, Oakes_82, balakrishnan2006generalized}. Other authors, as \cite{Aalen_88, Hougaard_94, Price_01, peng_2007, yu_2008, calsavara2013} considered cure rate models with a frailty term.

This manuscript proposes a new Bayesian cure rate model with a frailty term in risk function of latent competing causes, called power variance function frailty cure rate model (PVFCR). The proposed model is obtained of \cite{cancho2011flexible} models by adding a random effect (frailty term) on the baseline hazard function that acts multiplicatively in promotion time of each latent competitive cause. This approach allows that the competitive causes have different frailties, and that the most frail will fail earlier than the less frail. The distribution of the random effect is full based on family of power variance function (PVF) distributions suggested by \cite{tweedie1984index} and derived independently by \cite{hougaard1986a}. Besides, we consider that the number of competing causes related to the occurrence of an event of interest is modeled by the negative binomial distribution. Another advantage of the proposed model is that the negative binomial and PVF are flexible distributions and they include as particular cases well-known distributions, which can be tested for the best fitting in a straightforward way.

Our paper is organized as follows. In Section~\ref{secao2} we formulate the proposed model and Bayesian inference is described in Section~\ref{secao_inference}. In Section~\ref{secao_simulation} we consider a simulation study under different scenarios, where we numerically evaluate the performance of the Bayesian estimators as well as the performance of the proposed model in terms of Conditional Predictive Ordinate (CPO) criterion when it is compared to usual cure rate model \cite{cancho2011flexible}. An application to a real data set is presented in Section~\ref{aplicacao}. Finally, some final remarks are considered in Section~\ref{conclusoes}.

\section{Frailty cure rate model}\label{secao2}

The time for the $j$th competing cause to produce the event of interest (promotion time) is denoted by $Z_j$, $j=1,\ldots,N$, where $N$ represents the number of competing causes. The variable $N$ is unobservable with probability mass function (p.m.f) $p_n=P(N=n|\mbox{\boldmath{$\Theta$}})$ for $n=0, 1, \ldots$. We assume that, conditional on $N$ and on the parameters vector  $\mbox{\boldmath{$\varphi$}}$, $Z_j$'s are i.i.d. with cumulative distribution function $F(t|$\mbox{\boldmath{$\varphi$}}$)$ and survival function $S(t|$\mbox{\boldmath{$\varphi$}}$)=1-F(t|$\mbox{\boldmath{$\varphi$}}$)$. Also, we assume that $Z_1, Z_2,\ldots$ are independent from $N$.

The observable time of the occurrence of the event of interest is defined as $T=\min\{Z_0, Z_1, \ldots$, $Z_N\}$, where $P(Z_0=\infty)=1$, which leads to a cure rate $p_0$ of the population not susceptible to the event occurrence.

Under this setup, according to \cite{Rodrigues} the cure rate survival function of the random variable $T$, conditional to vector parameters $\mbox{\boldmath{$\vartheta$}}$, is given by
\begin{eqnarray}
S_{pop}(t|\mbox{\boldmath{$\vartheta$}})=P(T\geq t|\mbox{\boldmath{$\vartheta$}})=\sum_{n=0}^{\infty}P(N=n|\mbox{\boldmath{$\Theta$}})[S(t|\mbox{\boldmath{$\varphi$}})]^{n}=A_{N}[S(t|\mbox{\boldmath{$\varphi$}})], \label{eq1}
\end{eqnarray}
where ${A}_{N}[\cdot]$ is the probability generating function (p.g.f) of the random variable $N$, which converges when $s=S(t|\mbox{\boldmath{$\varphi$}})\in [0,1]$.

From now on we suppose that the number of competing causes, $N$, conditional to $\mbox{\boldmath{$\Theta$}}=(\eta,\theta)^\top$, follows a negative binomial distribution \cite{saha_2005} with p.m.f
\begin{eqnarray*}
	p_n=P(N=n|\mbox{\boldmath{$\Theta$}})=\frac{\Gamma(n+\eta^{-1})}{n!\Gamma(\eta^{-1})}\left(\frac{\eta \theta}{1+\eta \theta}\right)^{n}(1+\eta \theta)^{-1/\eta},\label{eq2}
\end{eqnarray*}
$n=0, 1, \ldots,$  $\theta>0$, $\eta\geq 0$ and $1+\eta\theta>0$, so that $E(N|\mbox{\boldmath{$\Theta$}})=\theta$ and $\mbox{Var}(N|\mbox{\boldmath{$\Theta$}})=\theta+\eta\theta^2$.

The p.g.f. is given by
\begin{eqnarray}
A_{N}(s)=\sum_{n=0}^{\infty}p_ns^n= \left\{1+\eta\theta(1-s)\right\}^{-1/\eta}, \quad 0\leq s \leq 1. \label{geradora}
\end{eqnarray}
As discussed by \cite{tournoud_2008}, the parameters of the negative binomial distribution have biological interpretations, which the mean number of competing causes is represented by $\theta$, whereas $\eta$ is the dispersion parameter.

So, taking into account $(\ref{geradora})$ in $(\ref{eq1})$, the population survival and density functions are given, respectively, by
\begin{eqnarray}
S_{pop}(t|\mbox{\boldmath{$\vartheta$}})=\{1+\eta\theta [1-S(t|\mbox{\boldmath{$\varphi$}})]\}^{-1/\eta}, \label{eq5}
\end{eqnarray}
and
\begin{eqnarray*}
	f_{pop}(t|\mbox{\boldmath{$\vartheta$}})=-\frac{d}{dt}S_{pop}(t|\mbox{\boldmath{$\varphi$}})=\theta f(t|\mbox{\boldmath{$\varphi$}})\Big\{1+\eta\theta \big[1-S(t|\mbox{\boldmath{$\varphi$}})\big]\Big\}^{-1/\eta-1},
	\label{eq6}
\end{eqnarray*}
where $f(t|\mbox{\boldmath{$\varphi$}})=-dF(t|\mbox{\boldmath{$\varphi$}})/dt$. The cure rate is determined by  $p_0=\lim_{t\to\infty} S_{pop}(t|\mbox{\boldmath{$\vartheta$}})=(1+\eta\theta)^{-1/\eta}>0$.

Usually, the most common choices for promotion time distribution that specify the function $S(t|\mbox{\boldmath{$\varphi$}})$ have been exponential, piecewise exponential, Weibull, among other. In order to capture the unobservable characteristics of each competing cause, we propose here to incorporate a random effect (frailty term) on the baseline hazard function that acts multiplicatively in promotion time. This approach allows that the competitive causes have different frailties, and that the most frail will fail earlier than the less frail \cite{wienke2010frailty}.

Let a nonnegative unobservable random variable $V$ that denote the frailty term. The hazard function of the $j$th competing cause is given by
\begin{eqnarray*}
	h(t|v_j,\mbox{\boldmath{$\varphi$}})=v_jh_0(t|\mbox{\boldmath{$\varphi$}}),\label{riscocondicional}
\end{eqnarray*}
where $v_j$ represents the frailty for the $j$th cause and $h_0(\cdot|\mbox{\boldmath{$\varphi$}})$ is baseline hazard function. 
The conditional survival function is easily obtained and it is given by
\begin{eqnarray} \displaystyle
S(t|v_j,\mbox{\boldmath{$\varphi$}})= S_0(t|\mbox{\boldmath{$\varphi$}})^{v_j},\nonumber
\end{eqnarray}
where $S_0(\cdot|\mbox{\boldmath{$\varphi$}})$ denotes the baseline survival function.

In this paper, we consider that the random variable $V$ follows the family of power variance function (PVF) distributions with parameters $\mu$, $\psi$ and $\gamma$, suggested by \cite{tweedie1984index} and derived independently by \cite{hougaard1986a}. For more PVF distribution details  \cite[see][]{wienke2010frailty}. We consider that $E(V|\mu,\psi,\gamma)=\mu=1$ and $\mbox{Var}(V|\mu,\psi,\gamma)=\mu^2/\psi=\sigma^2$, where $\sigma^2$ is interpreted as a measure of unobserved heterogeneity. With this restriction, the results PVF parameters are $\gamma$ and $\sigma^2$.

In order to eliminate the unobserved quantities, the random effect can be integrated out. Thus, marginal survival function is given by
\begin{eqnarray}
S(t|\mbox{\boldmath{$\varphi$}}^*)=E_{V}[S(t|v_j,\mbox{\boldmath{$\varphi$}})]=\int_{0}^{\infty} e^{-H_0(t|,\mbox{\boldmath{$\varphi$}})v_j}f_v(v_j|\gamma,\sigma^2)dv_j  =  \hspace{0.05cm} L_v[H_0(t|\mbox{\boldmath{$\varphi$}})],\nonumber\label{laplace}
\end{eqnarray}
where $\mbox{\boldmath{$\varphi$}}^*=(\mbox{\boldmath{$\varphi$}},\gamma,\sigma^2)^\top$, $f_v(\cdot|\gamma,\sigma^2)$ is the density function of $V$ conditional to $\gamma$ and $\sigma^2$, $H_0(\cdot|\mbox{\boldmath{$\varphi$}})$ is cumulative baseline hazard function and $L_v[\cdot]$ denotes the Laplace transform of frailty distribution.

The unconditional survival and density functions in the PVF frailty model is expressed by
\begin{eqnarray}
\displaystyle S(t|\mbox{\boldmath{$\varphi$}}^*)=\exp\left\{\frac{1-\gamma}{\gamma \sigma^2}\left[1-\left(1+\frac{\sigma^2 H_0(t|\mbox{\boldmath{$\varphi$}})}{1-\gamma}\right)^{\gamma}\right]\right\}\label{sobrevivencia_frag}
\end{eqnarray}
and
\begin{small}
	\begin{eqnarray}
	\displaystyle f(t|\mbox{\boldmath{$\varphi$}}^*)=h_0(t|\mbox{\boldmath{$\varphi$}})\left(1+\frac{\sigma^2H_0(t|\mbox{\boldmath{$\varphi$}})}{1-\gamma}\right)^{\gamma-1}\exp\left\{\frac{1-\gamma}{\gamma \sigma^2}\left[1-\left(1+\frac{\sigma^2 H_0(t|\mbox{\boldmath{$\varphi$}})}{1-\gamma}\right)^{\gamma}\right]\right\}.\label{densidade_frag}
	\end{eqnarray}
\end{small}

Besides providing an algebraic treatment of the closed-form for the marginal survival, the PVF family is a flexible model in the sense to include many other frailty models as special cases. For instance, the gamma frailty model is obtained if $\gamma=0$ and in the case of $\gamma=0.5$, the inverse Gaussian distribution is derived. The positive stable is a special case of the PVF distribution, however to show this fact, some asymptotic considerations are necessary. We refer the interested readers to \cite{wienke2010frailty}.

This way, as an alternative to the usual cure rate models (\ref{eq5}), we propose a new model that incorporates a frailty term for each competing cause and consider that, conditional on $N=n$ and on $\mbox{\boldmath{$\varphi$}}^*$, the latent times follow a survival function as in $(\ref{sobrevivencia_frag})$. As the number of competing causes follows a negative binomial distribution, the population survival function with PVF frailty is given by
\begin{eqnarray}
S_{pop}(t|\mbox{\boldmath{$\vartheta$}})=\left\{1+\eta\theta\left(1-\exp\left\{\frac{1-\gamma}{\gamma \sigma^2}\left[1-\left(1+\frac{\sigma^2 H_0(t|\mbox{\boldmath{$\varphi$}})}{1-\gamma}\right)^{\gamma}\right]\right\}\right)\right\}^{-1/\eta}, \label{sfragilidade1_1}
\end{eqnarray}
where $\mbox{\boldmath{$\vartheta$}}=(\mbox{\boldmath{$\varphi$}}^*,\mbox{\boldmath{$\Theta$}})^\top$.

We assume a Weibull distribution for the cumulative baseline hazard function, given by $H_0(t|\mbox{\boldmath{$\varphi$}})=e^{\alpha}t^{\lambda}$, where $\alpha\in \mathbb{R}$, $\lambda>0$ and $\mbox{\boldmath{$\varphi$}}=(\alpha,\lambda)^{\top}$.

Henceforward, we will refer to the model of which the survival function is as shown in $(\ref{sfragilidade1_1})$, by PVF frailty cure rate model or simply PVFCR model. Note that usual cure rate model (CR) $(\ref{eq5})$ is obtained as $\sigma^2 \to 0$.

\section{Bayesian inference}
\label{secao_inference}

Let us consider the situation when the time to event is not completely observed and it is subject to right censoring. For a given sample of size $m$, the observed time for $i$th unit is $W_i=\min\{T_i, C_i\}$, with $T_i=\min\{Z_{i0}, Z_{i1}, \ldots, Z_
{iN_i}\}$ and $C_i$ the censoring time, for $i=1,\ldots,m$. Let $\delta_i=\mathbb{I}(T_i \leq C_i)$, that is, $\delta_i=1$ if $W_i=T_i$ and $\delta_i=0$ otherwise.

We include covariate through the expected number of competing causes by $E(N_i|\mbox{\boldmath{$\Theta$}})=\theta_i=\exp\{\textbf{x}_i^{\top}\mbox{\boldmath{$\beta$}}\}$, $i=1, \ldots, m$, where $\mbox{\boldmath{$\beta$}}$ is a $k\times 1$ vector of regression coefficients. The observed data are represented by $\mathbf{D}=(m,\mbox{\boldmath{$w$}},\mbox{\boldmath{$\delta$}},\bf{X})$, $\mbox{\boldmath{$w$}}=(w_1, \ldots, w_m)^{\top}$, $\mbox{\boldmath{$\delta$}}=(\delta_1, \ldots, \delta_m)^{\top}$ and $\bf{X}$ is an $m\times k$ matrix containing the covariates.

The likelihood function of parameters
$\mbox{\boldmath{$\vartheta$}}=(\mbox{\boldmath{$\varphi$}}^*,\mbox{\boldmath{$\Theta$}})^\top=(\alpha,\lambda,\gamma,\sigma^2,\eta,\mbox{\boldmath{$\beta$}})^{\top}$ under non-informative censoring can be written as
\begin{eqnarray*}
	\displaystyle L(\mbox{\boldmath{$\vartheta$}}| \mathbf{D})\propto && \displaystyle \prod_{i=1}^{m} \{f_{pop}(w_i| \mbox{\boldmath{$\vartheta$}})\}^{\delta_i} \{S_{pop}(w_i| \mbox{\boldmath{$\vartheta$}})\}^{1-\delta_i} \nonumber \\
	\propto && \displaystyle \prod_{i=1}^{m} \Big\{\exp({\bf x}_i^{\top}{\mbox{\boldmath{$\beta$}}})f(w_i|\mbox{\boldmath{$\varphi$}}^*)\Big\}^{\delta_i} \Big\{1+\eta\exp({\bf x}_i^{\top}{\mbox{\boldmath{$\beta$}}})[1-S(w_i|\mbox{\boldmath{$\varphi$}}^*)]\Big\}^{-\frac{1}{\eta}-\delta_i},\label{verossimilhanca1}
\end{eqnarray*}
where $S(w_i|\mbox{\boldmath{$\varphi$}}^*)$ and $f(w_i|\mbox{\boldmath{$\varphi$}}^*)$ are given in (\ref{sobrevivencia_frag}) and (\ref{densidade_frag}), respectively.

The posterior distribution of $\mbox{\boldmath{$\vartheta$}}$ comes out to be
\begin{eqnarray}
\pi(\mbox{\boldmath{$\vartheta$}}| \mathbf{D}) &\propto& \pi(\mbox{\boldmath{$\vartheta$}})\lambda^{r} \exp\left\{\sum_{i=1}^{m}\delta_i x_i^{\top}{\mbox{\boldmath{$\beta$}}} + r\left(\alpha+\frac{1-\gamma}{\gamma \sigma^2}\right)\right\} \prod_{i=1}^{m}\left\{w_i^{\lambda-1}\left(1+\frac{\sigma^2e^{\alpha}w_i^{\lambda}}{1-\gamma}\right)^{\gamma-1}\right\}^{\delta_i} \nonumber \\
&\times& \prod_{i=1}^{m}\left\{1+\eta\exp(x_i^{\top}{\mbox{\boldmath{$\beta$}}}) \left[1-\exp\left\{\frac{1-\gamma}{\gamma \sigma^2}\left[1-\left(1+\frac{\sigma^2 e^{\alpha}w_i^{\lambda}}{1-\gamma}\right)^{\gamma}\right]\right\}\right]\right\}^{-1/\eta-\delta_i} \nonumber \\
&\times& \prod_{i=1}^{m}\exp\left\{-\left(\frac{1-\gamma}{\gamma\sigma^2}\right)\left(1+\frac{\sigma^2e^{\alpha}w_i^{\lambda}}{1-\gamma}\right)^{\gamma}\right\}^{\delta_i}, \label{posteriori}
\end{eqnarray}
where $r=\sum_{i=1}^{m}\delta_i$ and $\pi(\mbox{\boldmath{$\vartheta$}})$ is prior distribution of $\mbox{\boldmath{$\vartheta$}}$.

We consider independent prior distributions defining them as  $\mbox{\boldmath{$\beta$}} \sim \mbox{Normal}_{k+1}({\bf 0},100 {\bf I})$, $\alpha \sim \mbox{Normal}(0,100)$, $\gamma\sim \mbox{Uniform}(0,1)$ and $\eta$, $\lambda$ and $\sigma^2$ follow gamma distribution with mean $1$ for all and variances $1$, $100$ and $1$, respectively.

\subsection{Estimation procedure}

The posterior density of $\mbox{\boldmath{$\vartheta$}}$ in (\ref{posteriori}) is analytically intractable because the integration of the joint density is not easy to perform. An alternative is to rely on Markov chain Monte Carlo (MCMC) simulations. Here we consider Adaptive Metropolis Hasting algorithm with a multivariate distribution as proposal distribution \cite{Haario} implemented in the statistical package \textit{LaplacesDemon} \cite{laplacesDemon}, which provides a friendly environment for Bayesian inference within the R program \cite{R1}.

As a result, a sample of size $n_p$ from the joint posterior distribution of $\mbox{\boldmath{$\vartheta$}}$ is obtained (eliminating burn-in and jump samples). The sample from the posterior can be expressed as $(\mbox{\boldmath{$\vartheta$}}_{1},\mbox{\boldmath{$\vartheta$}}_{2},\ldots,\mbox{\boldmath{$\vartheta$}}_{n_p})$.
The estimator of $\mbox{\boldmath{$\vartheta$}}$ considered is given by
\begin{eqnarray}
\widehat{\mbox{\boldmath{$\vartheta$}}}=\frac{1}{n_p}\sum_{k=1}^{n_p}{\mbox{\boldmath{$\vartheta$}}_{k}}, \label{est_par_bayes}
\end{eqnarray}
and an estimator of the cure rate is
\begin{eqnarray}
\widehat{p}_0=\frac{1}{n_p}\sum_{k=1}^{n_p}{(1+\eta_k\theta_k)^{-1/\eta_k}}. \label{est_p0_bayes}
\end{eqnarray}

Consider the functions $Y_{k}(t)=S_{pop}(t|\mbox{\boldmath{$\vartheta$}}_{k})$ where $S_{pop}(t|\mbox{\boldmath{$\vartheta$}}_{k})$ is presented in (\ref{sfragilidade1_1}), conditional to $\mbox{\boldmath{$\vartheta$}}_{k}$. The proposed estimator of the improper survival function is
\begin{eqnarray}
\widehat{S_{pop}(t|\mbox{\boldmath{$\vartheta$}})}=\frac{1}{n_p}\sum_{k=1}^{n_p}{Y_{k}(t)},~~ \mbox{for each} ~ t>0.  \label{relia_bayes}
\end{eqnarray}

\subsection{Conditional predictive ordinate (CPO)}

A criterion for model selection that can be considered is based on the conditional predictive ordinates (CPO).

For an observed time to event ($\delta=1$), we define $g(t_i|\mbox{\boldmath{$\vartheta$}})=f_{pop}(t_i|\mbox{\boldmath{$\vartheta$}})$ and, for a censored time, $g(t_i|\mbox{\boldmath{$\vartheta$}})=S_{pop}(t_i|\mbox{\boldmath{$\vartheta$}})$. For the $i$th observation, $\mbox{CPO}_i$ can be expressed as
\begin{eqnarray}
\mbox{CPO}_i &=& \int g(t_i|\mbox{\boldmath{$\vartheta$}})\pi(\mbox{\boldmath{$\vartheta$}}|\mathcal{D}_{-i})d\mbox{\boldmath{$\vartheta$}} \nonumber \\
&=& \Bigg\{\int{\frac{\pi(\mbox{\boldmath{$\vartheta$}}|\mathcal{D})}{g(t_i|\mbox{\boldmath{$\vartheta$}})}d\mbox{\boldmath{$\vartheta$}}}\Bigg\}^{-1}. \nonumber
\end{eqnarray}

The $\mbox{CPO}_i$ can be interpreted as the height of the marginal density of the time to event at $t_i$. Thus, large values of $\mbox{CPO}_i$ imply a better fit of the model. For the proposed model, a closed form of the $\mbox{CPO}_i$ is not available. However, a Monte Carlo estimate of $\mbox{CPO}_i$ can be obtained by using a single MCMC sample from the posterior distribution $\pi(\mbox{\boldmath{$\vartheta$}}|\mathcal{D})$. A Monte Carlo approximation of $\mbox{CPO}_i$ is given by:
\begin{eqnarray}
\widehat{\mbox{CPO}}_i =  \Bigg\{\frac{1}{n_p}\sum_{k=1}^{n_p}{\frac{1}{g(t_i|\mbox{\boldmath{$\vartheta$}}_k)}}\Bigg\}^{-1}. \nonumber
\end{eqnarray}

A summary statistic of the $\mbox{CPO}_i$'s is the $\mbox{CPO}=\sum_{i=1}^m\log(\widehat{CPO}_i)$. The larger the value of CPO is, the better the fit of the model is \cite{rodrigues2012bayesian}.

\section{Simulation study}
\label{secao_simulation}

For data generation in this simulation study, we consider the model in (\ref{sfragilidade1_1}) with the Weibull distribution for the cumulative baseline hazard function with $\alpha=0$ and $\lambda=1$ (exponential distribution with rate $e^{\alpha}$) and one binary covariate $X$ values drawn from a Bernoulli distribution with parameter $0.5$.  We take for PVF frailty distribution $\gamma\in\{0.1, \ 0.5, \  0.9\}$ and $\sigma^2\in\{0.5, \ 1, \ 1.5, \ 2\}$.  The failure times data were simulated with $\eta=0.5$, $\theta_l=\exp(\beta_0+l\beta_1)$, $l=0, 1$, where $\beta_0=-0.5$ and $\beta_1=0.7$. In this way, $p_{0l}=(1+\eta\theta_l)^{-1/\eta}$, so that the cure rates for the two levels of $X$ are $p_{00}=0.59$ and $p_{01}=0.39$. The censoring times were sampled from the exponential distribution with $\tau$ parameter
(rate), where $\tau$ was set in order to control the proportion of censored observations. An algorithm to generate observed times and censoring indicators is:
\begin{enumerate}
	\item Draw $X_i\sim$ \mbox{Bernoulli}($0.5$) and $u_i\sim \mbox{Uniform}(0,1)$.
	\item Let $X_i=l$. If $u_i<p_{0l}$, $t_i=\infty$, otherwise, $$t_i=\frac{\left(1-\gamma\right)}{\sigma^2e^{\alpha}}\left(\left\{1-\frac{\gamma\sigma^2}{1-\gamma}\log{\left[1-\left(\frac{u^{-\eta}-1}{\eta\exp(\beta_0+\beta_1x_i)}\right)\right]}\right\}^{1/\gamma}-1\right).$$
	\item Draw $$c_i\sim \mbox{Exponential}\left(\tau\right), \quad \tau=\frac{e^{\eta}(p_{cl}-p_{0l})}{1-(p_{cl}-p_{0l})}, \quad \mbox{where} \quad p_{cl}=p_{0l}+0.01.$$
	\item Let $w_i=\min\{t_i, c_i\}$.
	\item If $t_i<c_i$, set $\delta_i=1$, otherwise, $\delta_i=0$, for $i=1, \ldots, m$.
\end{enumerate}

We consider four sample sizes, $m=100, 300, 500$ and $1000$. For each scenario (each combination of parameters values and sample size), we simulated $B=1000$ random samples.

As said previously, the Bayesian estimation procedures were performed using Adaptive Metropolis-Hastings algorithm such that the estimation of covariance matrix is update every $100$ iterations. For PVFCR and CR models, we generated $40000$ and $30000$ values for each parameter, respectively, disregarding the first $10000$ iterations to eliminate the effect of the initial values and spacing of size $30$ and $20$, respectively, to avoid correlation problems, obtaining a sample of size $n_p=1000$. The chains convergence was monitored for all simulation scenario, where good convergence results were obtained.

For each random sample, the estimates of $\mbox{\boldmath{$\vartheta$}}$ and cure rate are obtained by (\ref{est_par_bayes}) and (\ref{est_p0_bayes}). We computed the average of $B$ estimates of $\mbox{\boldmath{$\vartheta$}}$ (AE) and the root of the mean squared error (RMSE) of the estimators obtained from PVFCR and CR models. The results are all summarized in Tables \ref{gama_01}-\ref{gama_09}.

The results show that for both models, the average estimates of $p_{00}$ and $p_{01}$ were not affected by the increase of $\gamma$ and $\sigma^2$ values. Even for small sample sizes, the average estimates were close to fixed values. For the PVFCR model, we observe that the RMSEs appear reasonably close to zero as sample size increases, except for $\sigma^2$ parameter, which needs large sample size to close to zero. For a fixed sample size, the RMSE of $\sigma^2$ estimation increases as $\sigma^2$ also increases, regardless of $\gamma$ values.

We can note that the $\eta$ estimation obtained from CR model provides, in average, large RMSE, even when sample size is large, and this fact is more evident when $\gamma=0.1$ and $0.5$. However, if $\gamma=0.9$ the RMSE decreases as sample size increases.

It is worth mention that the inclusion of frailty term in the cure rate model (PVFCR) provides, in general, lower RMSE when compared to RMSE obtained by CR model. This behavior is clearly observed when $\gamma=0.1$ and $0.5$. Some exceptions occur, however for large sample size $(m=1000)$ the PVFCR model fit provides, in average, lower RMSE for the estimators, regardless of degree unobserved heterogeneity.

For models comparison, we considered the difference between the CPO values obtained under the fitted PVFCR and CR models. For a fixed scenario, we evaluate the mean difference and standard deviation of the $B=1000$ CPO's difference. This way, a positive CPO mean difference means that, in average, the CPO of the fitted PVFCR model is larger than CPO obtained from the fitted CR model, which shows advantage of the proposed model.

\begin{figure}[h] \centering
	\begin{center}
		\begin{minipage}[b]{0.41\linewidth}
			\includegraphics[width=\linewidth]{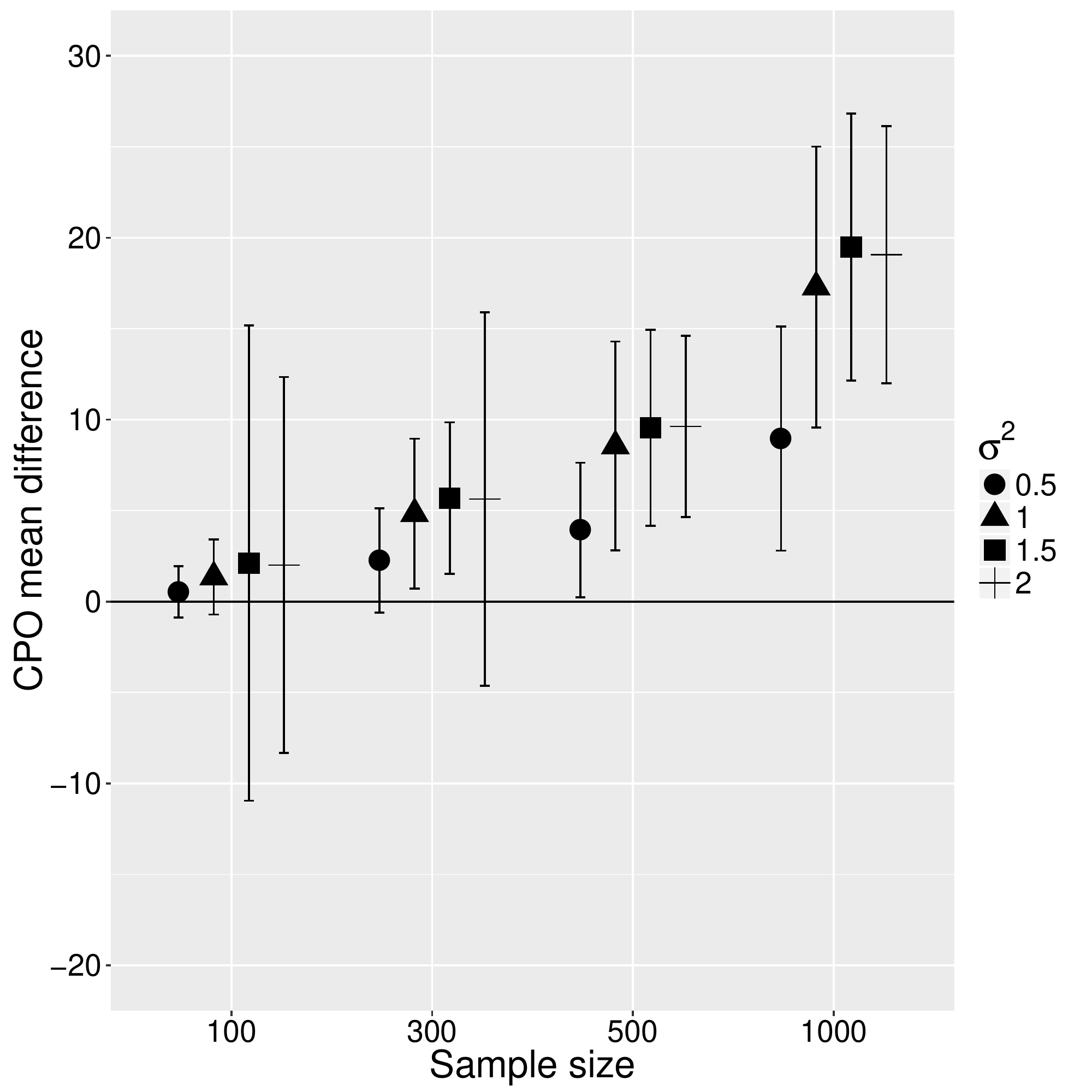}
			\subcaption{$\gamma=0.1$}
		\end{minipage}
		\begin{minipage}[b]{0.41\linewidth}
     	\includegraphics[width=\linewidth]{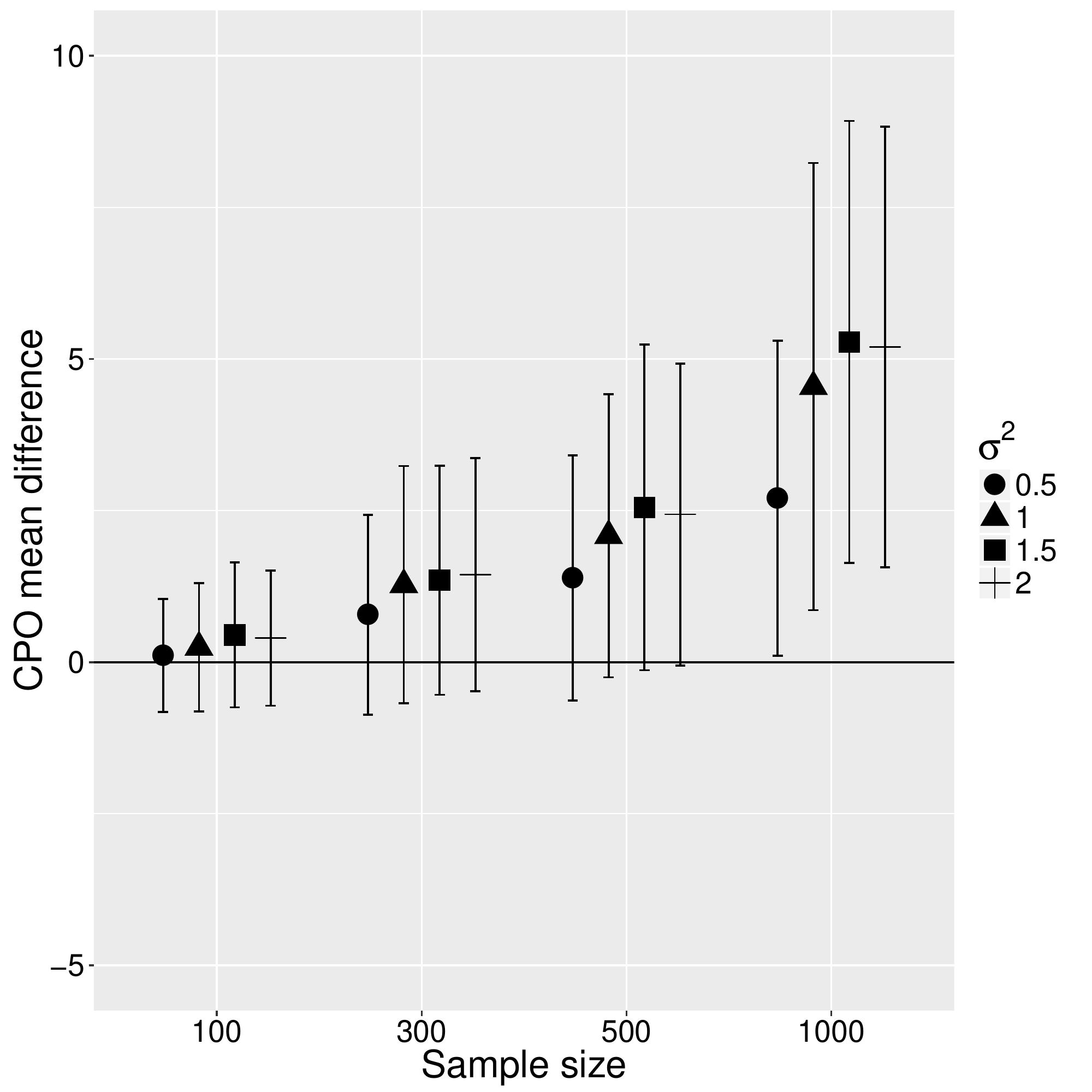}
	    \subcaption{$\gamma=0.5$}
       \end{minipage}
		\begin{minipage}[b]{0.41\linewidth}
     	\includegraphics[width=\linewidth]{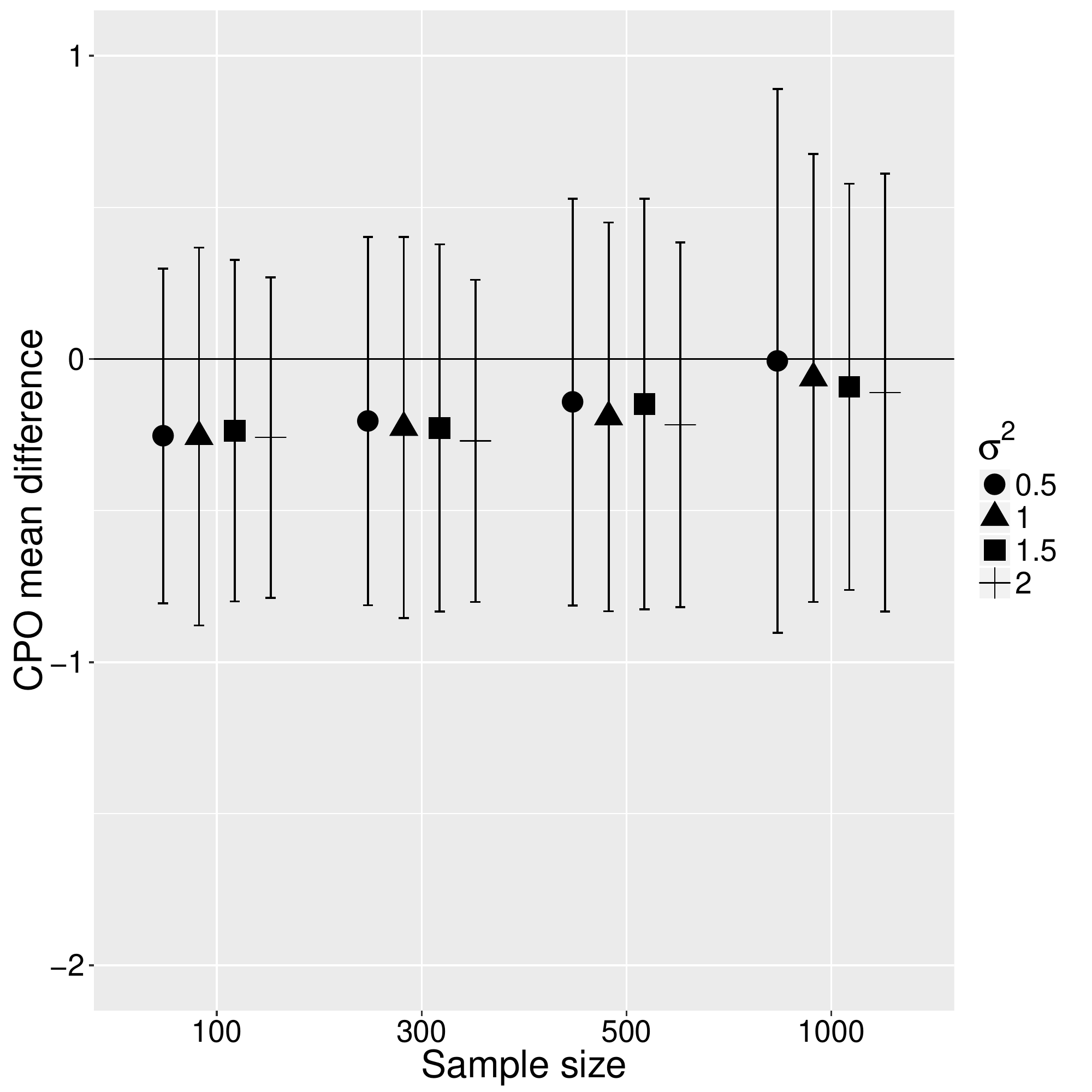}
    	\subcaption{$\gamma=0.9$}
        \end{minipage}   
		\caption{Mean difference (symbol) between the CPO values obtained under the fitted PVFCR and CR models and $\pm$ standard deviation of the difference (bar) when the data are generated from PVFCR model (a) for $\gamma=0.1$, (b)  for $\gamma=0.5$ and (c) for $\gamma=0.9$.}\label{fig_cpo}
	\end{center}
\end{figure}

In Figure~\ref{fig_cpo}, we present the CPO mean difference for all considered scenarios. For a fixed sample size and when $\gamma=0.1$ or $0.5$, CPO mean difference increases as $\sigma^2$ increases, which stabilizes in $\sigma^2=1.5$ and $2$. Besides, as sample size increases, CPO mean difference also increases, which indicates best fits of PVFCR model. By the other hand, when $\gamma=0.9$ the CPO mean difference is always negative, which favors the CR model, even with large unobserved heterogeneity.

\begin{table}[!h]
	\centering
 \begin{scriptsize}
  \caption{Root of the mean squared error (RMSE) and average of estimates (AE) of the estimators for simulated data from PVFCR model when $p_{00}=0.59$, $p_{01}=0.39$, $\beta_0=-0.5$, $\beta_1=0.7$, $\alpha=0$, $\lambda=1$, $\eta=0.5$ and $\gamma=0.1$.}
	{\begin{tabular}{ccc|cc|cc|cc|cc}
			\hline
			& \multicolumn{2}{c}{m} & \multicolumn{2}{c}{100} & \multicolumn{2}{c}{300} & \multicolumn{2}{c}{500} & \multicolumn{2}{c}{1000} \\
			\cline{2-11} $\sigma^2$          & Parameters & Model & RMSE & AE & RMSE & AE & RMSE & AE & RMSE & AE \\
			\hline
			\multicolumn{1}{c}{} & {$p_{00}$} & PVFCR & 0.073 & 0.586 & 0.040 & 0.585 & 0.032 & 0.588 & 0.023 & 0.589 \\
			\multicolumn{1}{c}{} &     & CR & 0.076 & 0.578 & 0.044 & 0.576 & 0.036 & 0.577 & 0.027 & 0.578 \\
			\multicolumn{1}{c}{} & {$p_{01}$} & PVFCR & 0.068 & 0.400 & 0.039 & 0.396 & 0.031 & 0.393 & 0.023 & 0.391 \\
			\multicolumn{1}{c}{} &     & CR & 0.066 & 0.396 & 0.039 & 0.395 & 0.031 & 0.393 & 0.024 & 0.394 \\
			\multicolumn{1}{c}{} &  {$\beta_{0}$} & PVFCR & 0.561 & -0.178 & 0.365 & -0.269 & 0.282 & -0.341 & 0.190 & -0.409 \\
			\multicolumn{1}{c}{} &     & CR & 0.763 & -0.013 & 0.609 & -0.058 & 0.532 & -0.102 & 0.448 & -0.142 \\
			\multicolumn{1}{c}{} &  {$\beta_{1}$} & PVFCR & 0.493 & 0.877 & 0.261 & 0.811 & 0.203 & 0.788 & 0.147 & 0.753 \\
			\multicolumn{1}{c}{} &     & CR & 0.515 & 0.904 & 0.302 & 0.869 & 0.258 & 0.867 & 0.215 & 0.858 \\
			\multicolumn{1}{c}{0.5} &  {$\eta$} & PVFCR & 1.075 & 1.428 & 0.871 & 1.175 & 0.696 & 0.988 & 0.503 & 0.795 \\
			\multicolumn{1}{c}{} &     & CR & 1.404 & 1.714 & 1.394 & 1.647 & 1.288 & 1.564 & 1.163 & 1.495 \\
			\multicolumn{1}{c}{} &  {$\alpha$} & PVFCR & 0.450 & -0.247 & 0.305 & -0.160 & 0.245 & -0.091 & 0.190 & -0.026 \\
			\multicolumn{1}{c}{} &     & CR & 0.784 & -0.687 & 0.700 & -0.640 & 0.656 & -0.611 & 0.617 & -0.589 \\
			\multicolumn{1}{c}{} &  {$\lambda$} & PVFCR & 0.203 & 1.103 & 0.126 & 1.057 & 0.100 & 1.046 & 0.080 & 1.039 \\
			\multicolumn{1}{c}{} &     & CR & 0.152 & 0.952 & 0.128 & 0.913 & 0.126 & 0.901 & 0.125 & 0.890 \\
			\multicolumn{1}{c}{} & {$\gamma$} & PVFCR & 0.352 & 0.439 & 0.320 & 0.398 & 0.292 & 0.368 & 0.237 & 0.312 \\
			\multicolumn{1}{c}{} &  {$\sigma^2$} & PVFCR & 0.642 & 1.065 & 0.532 & 0.965 & 0.491 & 0.917 & 0.413 & 0.821 \\
			\hline
			\multicolumn{1}{c}{} &  {$p_{00}$} & PVFCR & 0.072 & 0.586 & 0.043 & 0.586 & 0.032 & 0.587 & 0.022 & 0.589 \\
			\multicolumn{1}{c}{} &     & CR & 0.080 & 0.571 & 0.052 & 0.570 & 0.042 & 0.569 & 0.033 & 0.568 \\
			\multicolumn{1}{c}{} &  {$p_{01}$} & PVFCR & 0.070 & 0.400 & 0.039 & 0.393 & 0.031 & 0.395 & 0.022 & 0.391 \\
			\multicolumn{1}{c}{} &     & CR & 0.068 & 0.394 & 0.039 & 0.395 & 0.033 & 0.400 & 0.027 & 0.401 \\
			\multicolumn{1}{c}{} &  {$\beta_{0}$} & PVFCR & 0.578 & -0.144 & 0.396 & -0.251 & 0.316 & -0.310 & 0.217 & -0.384 \\
			\multicolumn{1}{c}{} &     & CR & 1.025 & 0.180 & 0.920 & 0.186 & 0.879 & 0.198 & 0.799 & 0.187 \\
			\multicolumn{1}{c}{} &  {$\beta_{1}$} & PVFCR & 0.498 & 0.890 & 0.286 & 0.840 & 0.211 & 0.797 & 0.153 & 0.772 \\
			\multicolumn{1}{c}{} &     & CR & 0.537 & 0.940 & 0.364 & 0.941 & 0.310 & 0.929 & 0.290 & 0.950 \\
			\multicolumn{1}{c}{{1}} &  {$\eta$} & PVFCR & 1.119 & 1.498 & 0.933 & 1.229 & 0.794 & 1.072 & 0.579 & 0.870 \\
			\multicolumn{1}{c}{} &     & CR & 1.730 & 2.021 & 1.938 & 2.163 & 2.008 & 2.244 & 1.968 & 2.285 \\
			\multicolumn{1}{c}{} &  {$\alpha$} & PVFCR & 0.602 & -0.409 & 0.420 & -0.240 & 0.335 & -0.153 & 0.240 & -0.074 \\
			\multicolumn{1}{c}{} &     & CR & 1.211 & -1.073 & 1.134 & -1.048 & 1.112 & -1.052 & 1.089 & -1.052 \\
			\multicolumn{1}{c}{} &  {$\lambda$} & PVFCR & 0.182 & 1.046 & 0.122 & 1.035 & 0.104 & 1.034 & 0.075 & 1.032 \\
			\multicolumn{1}{c}{} &     & CR & 0.192 & 0.858 & 0.188 & 0.838 & 0.190 & 0.832 & 0.180 & 0.832 \\
			\multicolumn{1}{c}{} &  {$\gamma$} & PVFCR & 0.298 & 0.377 & 0.224 & 0.293 & 0.182 & 0.252 & 0.138 & 0.213 \\
			\multicolumn{1}{c}{} &  {$\sigma^2$} & PVFCR & 0.547 & 1.324 & 0.533 & 1.329 & 0.529 & 1.336 & 0.443 & 1.287 \\
			\hline
			\multicolumn{1}{c}{} &  {$p_{00}$} & PVFCR & 0.077 & 0.581 & 0.044 & 0.586 & 0.032 & 0.586 & 0.024 & 0.587 \\
			\multicolumn{1}{c}{} &     & CR & 0.091 & 0.561 & 0.056 & 0.565 & 0.046 & 0.563 & 0.041 & 0.561 \\
			\multicolumn{1}{c}{} &  {$p_{01}$} & PVFCR & 0.069 & 0.401 & 0.042 & 0.397 & 0.035 & 0.396 & 0.024 & 0.391 \\
			\multicolumn{1}{c}{} &     & CR & 0.068 & 0.393 & 0.043 & 0.404 & 0.040 & 0.409 & 0.035 & 0.412 \\
			\multicolumn{1}{c}{} &  {$\beta_{0}$} & PVFCR & 0.713 & -0.067 & 0.422 & -0.236 & 0.345 & -0.286 & 0.234 & -0.369 \\
			\multicolumn{1}{c}{} &     & CR & 1.416 & 0.422 & 1.243 & 0.452 & 1.283 & 0.536 & 1.203 & 0.549 \\
			\multicolumn{1}{c}{} &  {$\beta_{1}$} & PVFCR & 0.508 & 0.888 & 0.277 & 0.819 & 0.211 & 0.798 & 0.158 & 0.771 \\
			\multicolumn{1}{c}{} &     & CR & 0.571 & 0.953 & 0.387 & 0.954 & 0.359 & 0.980 & 0.346 & 1.003 \\
			\multicolumn{1}{c}{1.5} &  {$\eta$} & PVFCR & 1.238 & 1.596 & 0.947 & 1.256 & 0.850 & 1.127 & 0.610 & 0.901 \\
			\multicolumn{1}{c}{} &     & CR & 2.010 & 2.273 & 2.544 & 2.704 & 2.811 & 2.969 & 2.809 & 3.089 \\
			\multicolumn{1}{c}{} &  {$\alpha$} & PVFCR & 0.845 & -0.594 & 0.480 & -0.291 & 0.397 & -0.224 & 0.279 & -0.127 \\
			\multicolumn{1}{c}{} &     & CR & 1.683 & -1.431 & 1.485 & -1.379 & 1.546 & -1.450 & 1.517 & -1.463 \\
			\multicolumn{1}{c}{} & {$\lambda$} & PVFCR & 0.176 & 0.993 & 0.123 & 1.015 & 0.103 & 1.020 & 0.073 & 1.017 \\
			\multicolumn{1}{c}{} &     & CR & 0.230 & 0.807 & 0.217 & 0.810 & 0.205 & 0.821 & 0.194 & 0.822 \\
			\multicolumn{1}{c}{} &  {$\gamma$} & PVFCR & 0.277 & 0.357 & 0.182 & 0.251 & 0.142 & 0.215 & 0.099 & 0.176 \\
			\multicolumn{1}{c}{} &  {$\sigma^2$} & PVFCR & 0.521 & 1.484 & 0.611 & 1.699 & 0.590 & 1.754 & 0.480 & 1.724 \\
			\hline
			\multicolumn{1}{c}{} &  {$p_{00}$} & PVFCR & 0.082 & 0.586 & 0.044 & 0.585 & 0.034 & 0.589 & 0.025 & 0.587 \\
			\multicolumn{1}{c}{} &     & CR & 0.102 & 0.560 & 0.061 & 0.562 & 0.049 & 0.564 & 0.044 & 0.559 \\
			\multicolumn{1}{c}{} &  {$p_{01}$} & PVFCR & 0.078 & 0.408 & 0.043 & 0.400 & 0.035 & 0.398 & 0.024 & 0.392 \\
			\multicolumn{1}{c}{} &     & CR & 0.079 & 0.396 & 0.045 & 0.410 & 0.045 & 0.418 & 0.043 & 0.423 \\
			\multicolumn{1}{c}{} &  {$\beta_{0}$} & PVFCR & 1.292 & -0.052 & 0.433 & -0.216 & 0.310 & -0.311 & 0.244 & -0.361 \\
			\multicolumn{1}{c}{} &     & CR & 2.522 & 0.621 & 1.758 & 0.725 & 1.544 & 0.749 & 1.620 & 0.902 \\
			\multicolumn{1}{c}{} &  {$\beta_{1}$} & PVFCR & 0.580 & 0.879 & 0.280 & 0.820 & 0.215 & 0.795 & 0.159 & 0.773 \\
			\multicolumn{1}{c}{} &     & CR & 0.760 & 0.962 & 0.418 & 0.984 & 0.392 & 1.009 & 0.393 & 1.051 \\
			\multicolumn{1}{c}{2} &  {$\eta$} & PVFCR & 1.620 & 1.610 & 1.005 & 1.310 & 0.777 & 1.090 & 0.631 & 0.919 \\
			\multicolumn{1}{c}{} &     & CR & 3.092 & 2.435 & 3.134 & 3.191 & 3.314 & 3.462 & 3.658 & 3.872 \\
			\multicolumn{1}{c}{} &  {$\alpha$} & PVFCR & 1.231 & -0.713 & 0.567 & -0.397 & 0.425 & -0.251 & 0.333 & -0.171 \\
			\multicolumn{1}{c}{} &     & CR & 2.424 & -1.705 & 1.972 & -1.711 & 1.840 & -1.723 & 1.924 & -1.839 \\
			\multicolumn{1}{c}{} &  {$\lambda$} & PVFCR & 0.201 & 0.938 & 0.123 & 0.977 & 0.103 & 0.997 & 0.077 & 1.003 \\
			\multicolumn{1}{c}{} &     & CR & 0.320 & 0.770 & 0.234 & 0.796 & 0.219 & 0.808 & 0.196 & 0.830 \\
			\multicolumn{1}{c}{} &  {$\gamma$} & PVFCR & 0.275 & 0.357 & 0.171 & 0.241 & 0.120 & 0.194 & 0.085 & 0.162 \\
			\multicolumn{1}{c}{} &  {$\sigma^2$} & PVFCR & 0.711 & 1.551 & 0.670 & 1.970 & 0.642 & 2.110 & 0.609 & 2.192 \\
			\hline
	  \end{tabular}}
	\label{gama_01}
	\end{scriptsize}
\end{table}

\begin{table}[!h]
	\centering
	 \begin{scriptsize}
	\caption{Root of the mean squared error (RMSE) and average of estimates (AE) of the estimators for simulated data from PVFCR model when $p_{00}=0.59$, $p_{01}=0.39$, $\beta_0=-0.5$, $\beta_1=0.7$, $\alpha=0$, $\lambda=1$, $\eta=0.5$ and $\gamma=0.5$.}
	{\begin{tabular}{ccc|cc|cc|cc|cc}
			\hline
		& \multicolumn{2}{c}{m} & \multicolumn{2}{c}{100} & \multicolumn{2}{c}{300} & \multicolumn{2}{c}{500} & \multicolumn{2}{c}{1000} \\
			\cline{2-11}  $\sigma^2$         & Parameters & Model & RMSE & AE  & RMSE & AE  & RMSE & AE  & RMSE & AE \\
			\hline
			\multicolumn{1}{c}{} & {$p_{00}$} & PVFCR & 0.071 & 0.587 & 0.039 & 0.588 & 0.032 & 0.587 & 0.022 & 0.588 \\
			\multicolumn{1}{c}{} &     & CR  & 0.073 & 0.580 & 0.042 & 0.581 & 0.035 & 0.581 & 0.024 & 0.581 \\
			\multicolumn{1}{c}{} & {$p_{01}$} & PVFCR & 0.069 & 0.403 & 0.038 & 0.395 & 0.031 & 0.392 & 0.022 & 0.389 \\
			\multicolumn{1}{c}{} &     & CR  & 0.067 & 0.399 & 0.037 & 0.394 & 0.031 & 0.393 & 0.022 & 0.391 \\
			\multicolumn{1}{c}{} & {$\beta_{0}$} & PVFCR & 0.545 & -0.192 & 0.338 & -0.299 & 0.294 & -0.334 & 0.198 & -0.393 \\
			\multicolumn{1}{c}{} &     & CR  & 0.696 & -0.067 & 0.505 & -0.152 & 0.473 & -0.172 & 0.367 & -0.219 \\
			\multicolumn{1}{c}{} & {$\beta_{1}$} & PVFCR & 0.465 & 0.857 & 0.257 & 0.810 & 0.203 & 0.793 & 0.150 & 0.765 \\
			\multicolumn{1}{c}{} &     & CR  & 0.484 & 0.883 & 0.291 & 0.857 & 0.246 & 0.852 & 0.204 & 0.842 \\
			\multicolumn{1}{c}{0.5} & {$\eta$} & PVFCR & 1.055 & 1.394 & 0.825 & 1.116 & 0.727 & 1.008 & 0.531 & 0.835 \\
			\multicolumn{1}{c}{} &     & CR  & 1.322 & 1.619 & 1.197 & 1.453 & 1.153 & 1.409 & 0.981 & 1.303 \\
			\multicolumn{1}{c}{} & {$\alpha$} & PVFCR & 0.405 & -0.210 & 0.289 & -0.155 & 0.247 & -0.133 & 0.188 & -0.092 \\
			\multicolumn{1}{c}{} &     & CR  & 0.664 & -0.572 & 0.566 & -0.509 & 0.543 & -0.496 & 0.495 & -0.467 \\
			\multicolumn{1}{c}{} & {$\lambda$} & PVFCR & 0.205 & 1.116 & 0.116 & 1.053 & 0.095 & 1.039 & 0.065 & 1.019 \\
			\multicolumn{1}{c}{} &     & CR  & 0.142 & 0.991 & 0.100 & 0.951 & 0.094 & 0.943 & 0.089 & 0.930 \\
			\multicolumn{1}{c}{} & {$\gamma$} & PVFCR & 0.085 & 0.471 & 0.103 & 0.481 & 0.110 & 0.476 & 0.112 & 0.467 \\
			\multicolumn{1}{c}{} & {$\sigma^2$} & PVFCR & 0.545 & 0.988 & 0.445 & 0.891 & 0.407 & 0.844 & 0.381 & 0.788 \\
			\hline
			\multicolumn{1}{c}{} & {$p_{00}$} & PVFCR & 0.073 & 0.588 & 0.041 & 0.588 & 0.033 & 0.588 & 0.023 & 0.587 \\
			\multicolumn{1}{c}{} &     & CR  & 0.075 & 0.581 & 0.044 & 0.580 & 0.036 & 0.579 & 0.028 & 0.577 \\
			\multicolumn{1}{c}{} & {$p_{01}$} & PVFCR & 0.072 & 0.404 & 0.038 & 0.396 & 0.031 & 0.394 & 0.021 & 0.390 \\
			\multicolumn{1}{c}{} &     & CR  & 0.070 & 0.401 & 0.038 & 0.395 & 0.031 & 0.395 & 0.023 & 0.394 \\
			\multicolumn{1}{c}{} & {$\beta_{0}$} & PVFCR & 0.560 & -0.182 & 0.392 & -0.262 & 0.318 & -0.307 & 0.232 & -0.367 \\
			\multicolumn{1}{c}{} &     & CR  & 0.738 & -0.032 & 0.631 & -0.053 & 0.572 & -0.074 & 0.505 & -0.091 \\
			\multicolumn{1}{c}{} & {$\beta_{1}$} & PVFCR & 0.481 & 0.871 & 0.288 & 0.837 & 0.218 & 0.810 & 0.159 & 0.775 \\
			\multicolumn{1}{c}{} &     & CR  & 0.507 & 0.904 & 0.334 & 0.898 & 0.277 & 0.889 & 0.236 & 0.882 \\
			\multicolumn{1}{c}{1} & {$\eta$} & PVFCR & 1.097 & 1.444 & 0.948 & 1.227 & 0.805 & 1.095 & 0.618 & 0.905 \\
			\multicolumn{1}{c}{} &     & CR  & 1.414 & 1.716 & 1.472 & 1.703 & 1.398 & 1.663 & 1.307 & 1.626 \\
			\multicolumn{1}{c}{} & {$\alpha$} & PVFCR & 0.538 & -0.375 & 0.419 & -0.310 & 0.360 & -0.267 & 0.290 & -0.192 \\
			\multicolumn{1}{c}{} &     & CR  & 0.858 & -0.769 & 0.802 & -0.746 & 0.768 & -0.725 & 0.741 & -0.713 \\
			\multicolumn{1}{c}{} & {$\lambda$} & PVFCR & 0.168 & 1.064 & 0.108 & 1.015 & 0.086 & 1.001 & 0.069 & 0.996 \\
			\multicolumn{1}{c}{} &     & CR  & 0.145 & 0.935 & 0.135 & 0.901 & 0.131 & 0.894 & 0.127 & 0.888 \\
			\multicolumn{1}{c}{} & {$\gamma$} & PVFCR & 0.094 & 0.455 & 0.121 & 0.444 & 0.124 & 0.436 & 0.127 & 0.422 \\
			\multicolumn{1}{c}{} & {$\sigma^2$} & PVFCR & 0.293 & 1.056 & 0.278 & 0.990 & 0.273 & 0.956 & 0.325 & 0.960 \\
			\hline
			\multicolumn{1}{c}{} & {$p_{00}$} & PVFCR & 0.068 & 0.585 & 0.042 & 0.585 & 0.033 & 0.586 & 0.023 & 0.586 \\
			\multicolumn{1}{c}{} &     & CR  & 0.071 & 0.576 & 0.045 & 0.576 & 0.037 & 0.576 & 0.029 & 0.574 \\
			\multicolumn{1}{c}{} & {$p_{01}$} & PVFCR & 0.068 & 0.399 & 0.039 & 0.395 & 0.031 & 0.394 & 0.022 & 0.389 \\
			\multicolumn{1}{c}{} &     & CR  & 0.066 & 0.396 & 0.039 & 0.396 & 0.032 & 0.397 & 0.024 & 0.395 \\
			\multicolumn{1}{c}{} & {$\beta_{0}$} & PVFCR & 0.584 & -0.135 & 0.438 & -0.220 & 0.333 & -0.289 & 0.254 & -0.344 \\
			\multicolumn{1}{c}{} &     & CR  & 0.808 & 0.049 & 0.729 & 0.033 & 0.657 & 0.009 & 0.615 & 0.005 \\
			\multicolumn{1}{c}{} & {$\beta_{1}$} & PVFCR & 0.505 & 0.892 & 0.298 & 0.845 & 0.215 & 0.808 & 0.163 & 0.785 \\
			\multicolumn{1}{c}{} &     & CR  & 0.538 & 0.928 & 0.353 & 0.916 & 0.288 & 0.902 & 0.263 & 0.914 \\
			\multicolumn{1}{c}{1.5} & {$\eta$} & PVFCR & 1.167 & 1.516 & 1.048 & 1.309 & 0.836 & 1.130 & 0.673 & 0.961 \\
			\multicolumn{1}{c}{} &     & CR  & 1.548 & 1.850 & 1.660 & 1.878 & 1.590 & 1.847 & 1.554 & 1.856 \\
			\multicolumn{1}{c}{} & {$\alpha$} & PVFCR & 0.667 & -0.534 & 0.536 & -0.434 & 0.442 & -0.359 & 0.369 & -0.293 \\
			\multicolumn{1}{c}{} &     & CR  & 1.040 & -0.962 & 0.972 & -0.912 & 0.938 & -0.896 & 0.933 & -0.902 \\
			\multicolumn{1}{c}{} & {$\lambda$} & PVFCR & 0.164 & 1.027 & 0.108 & 0.983 & 0.089 & 0.973 & 0.073 & 0.970 \\
			\multicolumn{1}{c}{} &     & CR  & 0.164 & 0.892 & 0.159 & 0.872 & 0.156 & 0.861 & 0.154 & 0.860 \\
			\multicolumn{1}{c}{} & {$\gamma$} & PVFCR & 0.108 & 0.437 & 0.124 & 0.431 & 0.133 & 0.416 & 0.137 & 0.404 \\
			\multicolumn{1}{c}{} & {$\sigma^2$} & PVFCR & 0.496 & 1.118 & 0.538 & 1.064 & 0.533 & 1.080 & 0.540 & 1.095 \\
			\hline
			\multicolumn{1}{c}{} & {$p_{00}$} & PVFCR & 0.072 & 0.582 & 0.042 & 0.585 & 0.034 & 0.583 & 0.024 & 0.585 \\
			\multicolumn{1}{c}{} &     & CR  & 0.076 & 0.574 & 0.046 & 0.576 & 0.040 & 0.572 & 0.030 & 0.572 \\
			\multicolumn{1}{c}{} & {$p_{01}$} & PVFCR & 0.066 & 0.395 & 0.038 & 0.393 & 0.030 & 0.392 & 0.021 & 0.390 \\
			\multicolumn{1}{c}{} &     & CR  & 0.065 & 0.393 & 0.038 & 0.394 & 0.032 & 0.396 & 0.024 & 0.397 \\
			\multicolumn{1}{c}{} & {$\beta_{0}$} & PVFCR & 0.589 & -0.128 & 0.463 & -0.216 & 0.391 & -0.245 & 0.269 & -0.334 \\
			\multicolumn{1}{c}{} &     & CR  & 0.825 & 0.062 & 0.768 & 0.056 & 0.756 & 0.086 & 0.668 & 0.059 \\
			\multicolumn{1}{c}{} & {$\beta_{1}$} & PVFCR & 0.483 & 0.894 & 0.289 & 0.850 & 0.239 & 0.823 & 0.162 & 0.784 \\
			\multicolumn{1}{c}{} &     & CR  & 0.516 & 0.931 & 0.354 & 0.929 & 0.314 & 0.922 & 0.273 & 0.925 \\
			\multicolumn{1}{c}{2} & {$\eta$} & PVFCR & 1.140 & 1.502 & 1.056 & 1.307 & 0.961 & 1.214 & 0.696 & 0.979 \\
			\multicolumn{1}{c}{} &     & CR  & 1.529 & 1.845 & 1.727 & 1.923 & 1.777 & 1.996 & 1.676 & 1.981 \\
			\multicolumn{1}{c}{} & {$\alpha$} & PVFCR & 0.751 & -0.632 & 0.621 & -0.530 & 0.572 & -0.491 & 0.441 & -0.370 \\
			\multicolumn{1}{c}{} &     & CR  & 1.125 & -1.053 & 1.089 & -1.027 & 1.094 & -1.042 & 1.055 & -1.024 \\
			\multicolumn{1}{c}{} & {$\lambda$} & PVFCR & 0.154 & 0.988 & 0.110 & 0.957 & 0.099 & 0.950 & 0.083 & 0.950 \\
			\multicolumn{1}{c}{} &     & CR  & 0.182 & 0.864 & 0.181 & 0.845 & 0.174 & 0.847 & 0.171 & 0.842 \\
			\multicolumn{1}{c}{} & {$\gamma$} & PVFCR & 0.105 & 0.438 & 0.130 & 0.419 & 0.133 & 0.415 & 0.132 & 0.407 \\
			\multicolumn{1}{c}{} & {$\sigma^2$} & PVFCR & 0.920 & 1.147 & 0.948 & 1.112 & 0.946 & 1.117 & 0.890 & 1.211 \\
			\hline
	\end{tabular}}%
	\label{gama_05}%
	 \end{scriptsize}
\end{table}%

\begin{table}[h!]
	\centering
	 \begin{scriptsize}
	\caption{Root of the mean squared error (RMSE) and average of estimates (AE) of the estimators for simulated data from PVFCR model when $p_{00}=0.59$, $p_{01}=0.39$, $\beta_0=-0.5$, $\beta_1=0.7$, $\alpha=0$, $\lambda=1$, $\eta=0.5$ and $\gamma=0.9$.}
	\begin{tabular}{ccc|cc|cc|cc|cc}   
			\hline	
			& \multicolumn{2}{c}{m} & \multicolumn{2}{c}{100} & \multicolumn{2}{c}{300} & \multicolumn{2}{c}{500} & \multicolumn{2}{c}{1000} \\
			\cline{2-11}  $\sigma^2$         & Parameters & Model & RMSE & AE  & RMSE & AE  & RMSE & AE  & RMSE & AE \\
			\hline
			\multicolumn{1}{c}{} & {$p_{00}$} & PVFCR & 0.069 & 0.591 & 0.041 & 0.590 & 0.030 & 0.591 & 0.021 & 0.591 \\
			\multicolumn{1}{c}{} &     & CR  & 0.070 & 0.587 & 0.042 & 0.586 & 0.030 & 0.588 & 0.021 & 0.588 \\
			\multicolumn{1}{c}{} & {$p_{01}$} & PVFCR & 0.069 & 0.405 & 0.041 & 0.398 & 0.032 & 0.393 & 0.022 & 0.390 \\
			\multicolumn{1}{c}{} &     & CR  & 0.067 & 0.402 & 0.041 & 0.397 & 0.031 & 0.393 & 0.022 & 0.389 \\
			\multicolumn{1}{c}{} & {$\beta_{0}$} & PVFCR & 0.478 & -0.243 & 0.305 & -0.339 & 0.229 & -0.393 & 0.157 & -0.445 \\
			\multicolumn{1}{c}{} &     & CR  & 0.552 & -0.173 & 0.384 & -0.265 & 0.302 & -0.326 & 0.214 & -0.384 \\
			\multicolumn{1}{c}{} & {$\beta_{1}$} & PVFCR & 0.469 & 0.854 & 0.260 & 0.791 & 0.198 & 0.775 & 0.133 & 0.744 \\
			\multicolumn{1}{c}{} &     & CR  & 0.482 & 0.873 & 0.280 & 0.819 & 0.219 & 0.805 & 0.154 & 0.776 \\
			\multicolumn{1}{c}{0.5} & {$\eta$} & PVFCR & 0.968 & 1.321 & 0.736 & 1.027 & 0.600 & 0.875 & 0.416 & 0.704 \\
			\multicolumn{1}{c}{} &     & CR  & 1.123 & 1.456 & 0.937 & 1.204 & 0.795 & 1.049 & 0.583 & 0.876 \\
			\multicolumn{1}{c}{} & {$\alpha$} & PVFCR & 0.330 & -0.110 & 0.207 & -0.075 & 0.166 & -0.054 & 0.124 & -0.030 \\
			\multicolumn{1}{c}{} &     & CR  & 0.467 & -0.378 & 0.355 & -0.295 & 0.303 & -0.255 & 0.247 & -0.216 \\
			\multicolumn{1}{c}{} & {$\lambda$} & PVFCR & 0.235 & 1.162 & 0.128 & 1.084 & 0.097 & 1.058 & 0.062 & 1.031 \\
			\multicolumn{1}{c}{} &     & CR  & 0.156 & 1.062 & 0.091 & 1.017 & 0.073 & 1.000 & 0.054 & 0.982 \\
			\multicolumn{1}{c}{} & {$\gamma$} & PVFCR & 0.394 & 0.511 & 0.335 & 0.572 & 0.307 & 0.602 & 0.278 & 0.634 \\
			\multicolumn{1}{c}{} & {$\sigma^2$} & PVFCR & 0.437 & 0.901 & 0.345 & 0.813 & 0.326 & 0.785 & 0.325 & 0.764 \\
			\hline
			\multicolumn{1}{c}{} & {$p_{00}$} & PVFCR & 0.072 & 0.593 & 0.039 & 0.591 & 0.030 & 0.589 & 0.022 & 0.589 \\
			\multicolumn{1}{c}{} &     & CR  & 0.072 & 0.589 & 0.039 & 0.587 & 0.031 & 0.586 & 0.023 & 0.587 \\
			\multicolumn{1}{c}{} & {$p_{01}$} & PVFCR & 0.066 & 0.400 & 0.041 & 0.396 & 0.031 & 0.393 & 0.021 & 0.390 \\
			\multicolumn{1}{c}{} &     & CR  & 0.065 & 0.397 & 0.040 & 0.395 & 0.031 & 0.392 & 0.021 & 0.389 \\
			\multicolumn{1}{c}{} & {$\beta_{0}$} & PVFCR & 0.490 & -0.253 & 0.296 & -0.350 & 0.229 & -0.382 & 0.164 & -0.439 \\
			\multicolumn{1}{c}{} &     & CR  & 0.570 & -0.180 & 0.376 & -0.279 & 0.298 & -0.317 & 0.220 & -0.378 \\
			\multicolumn{1}{c}{} & {$\beta_{1}$} & PVFCR & 0.478 & 0.885 & 0.262 & 0.799 & 0.196 & 0.768 & 0.134 & 0.739 \\
			\multicolumn{1}{c}{} &     & CR  & 0.495 & 0.906 & 0.281 & 0.826 & 0.215 & 0.798 & 0.154 & 0.770 \\
			\multicolumn{1}{c}{1} & {$\eta$} & PVFCR & 0.979 & 1.322 & 0.729 & 1.007 & 0.587 & 0.883 & 0.422 & 0.702 \\
			\multicolumn{1}{c}{} &     & CR  & 1.144 & 1.460 & 0.922 & 1.175 & 0.770 & 1.052 & 0.589 & 0.874 \\
			\multicolumn{1}{c}{} & {$\alpha$} & PVFCR & 0.356 & -0.157 & 0.228 & -0.124 & 0.189 & -0.111 & 0.144 & -0.081 \\
			\multicolumn{1}{c}{} &     & CR  & 0.512 & -0.425 & 0.390 & -0.336 & 0.345 & -0.307 & 0.291 & -0.263 \\
			\multicolumn{1}{c}{} & {$\lambda$} & PVFCR & 0.226 & 1.152 & 0.117 & 1.069 & 0.085 & 1.045 & 0.055 & 1.016 \\
			\multicolumn{1}{c}{} &     & CR  & 0.150 & 1.053 & 0.087 & 1.004 & 0.069 & 0.989 & 0.059 & 0.969 \\
			\multicolumn{1}{c}{} & {$\gamma$} & PVFCR & 0.393 & 0.512 & 0.331 & 0.576 & 0.303 & 0.605 & 0.274 & 0.637 \\
			\multicolumn{1}{c}{} & {$\sigma^2$} & PVFCR & 0.193 & 0.908 & 0.233 & 0.813 & 0.260 & 0.790 & 0.293 & 0.772 \\
			\hline
			\multicolumn{1}{c}{} & {$p_{00}$} & PVFCR & 0.069 & 0.593 & 0.040 & 0.590 & 0.032 & 0.590 & 0.021 & 0.589 \\
			\multicolumn{1}{c}{} &     & CR  & 0.069 & 0.589 & 0.041 & 0.587 & 0.032 & 0.587 & 0.022 & 0.586 \\
			\multicolumn{1}{c}{} & {$p_{01}$} & PVFCR & 0.067 & 0.400 & 0.039 & 0.396 & 0.030 & 0.394 & 0.020 & 0.389 \\
			\multicolumn{1}{c}{} &     & CR  & 0.066 & 0.397 & 0.039 & 0.395 & 0.030 & 0.393 & 0.020 & 0.389 \\
			\multicolumn{1}{c}{} & {$\beta_{0}$} & PVFCR & 0.470 & -0.256 & 0.302 & -0.344 & 0.241 & -0.385 & 0.164 & -0.434 \\
			\multicolumn{1}{c}{} &     & CR  & 0.546 & -0.183 & 0.380 & -0.272 & 0.318 & -0.313 & 0.222 & -0.373 \\
			\multicolumn{1}{c}{} & {$\beta_{1}$} & PVFCR & 0.492 & 0.890 & 0.262 & 0.798 & 0.197 & 0.769 & 0.138 & 0.742 \\
			\multicolumn{1}{c}{} &     & CR  & 0.510 & 0.910 & 0.280 & 0.825 & 0.218 & 0.801 & 0.159 & 0.773 \\
			\multicolumn{1}{c}{1.5} & {$\eta$} & PVFCR & 0.991 & 1.321 & 0.720 & 1.014 & 0.607 & 0.884 & 0.441 & 0.716 \\
			\multicolumn{1}{c}{} &     & CR  & 1.148 & 1.460 & 0.910 & 1.185 & 0.812 & 1.069 & 0.606 & 0.888 \\
			\multicolumn{1}{c}{} & {$\alpha$} & PVFCR & 0.358 & -0.190 & 0.255 & -0.168 & 0.215 & -0.146 & 0.175 & -0.122 \\
			\multicolumn{1}{c}{} &     & CR  & 0.528 & -0.453 & 0.426 & -0.379 & 0.387 & -0.347 & 0.330 & -0.302 \\
			\multicolumn{1}{c}{} & {$\lambda$} & PVFCR & 0.212 & 1.137 & 0.108 & 1.063 & 0.082 & 1.038 & 0.055 & 1.010 \\
			\multicolumn{1}{c}{} &     & CR  & 0.144 & 1.040 & 0.081 & 0.999 & 0.072 & 0.982 & 0.063 & 0.964 \\
			\multicolumn{1}{c}{} & {$\gamma$} & PVFCR & 0.392 & 0.512 & 0.330 & 0.576 & 0.306 & 0.603 & 0.265 & 0.646 \\
			\multicolumn{1}{c}{} & {$\sigma^2$} & PVFCR & 0.615 & 0.908 & 0.696 & 0.819 & 0.719 & 0.797 & 0.742 & 0.779 \\
			\hline
			\multicolumn{1}{c}{} & {$p_{00}$} & PVFCR & 0.069 & 0.596 & 0.040 & 0.590 & 0.030 & 0.590 & 0.021 & 0.589 \\
			\multicolumn{1}{c}{} &     & CR  & 0.069 & 0.592 & 0.041 & 0.587 & 0.030 & 0.587 & 0.022 & 0.586 \\
			\multicolumn{1}{c}{} & {$p_{01}$} & PVFCR & 0.069 & 0.401 & 0.039 & 0.398 & 0.032 & 0.394 & 0.022 & 0.391 \\
			\multicolumn{1}{c}{} &     & CR  & 0.068 & 0.398 & 0.038 & 0.396 & 0.032 & 0.393 & 0.022 & 0.391 \\
			\multicolumn{1}{c}{} & {$\beta_{0}$} & PVFCR & 0.477 & -0.264 & 0.296 & -0.347 & 0.225 & -0.390 & 0.167 & -0.432 \\
			\multicolumn{1}{c}{} &     & CR  & 0.554 & -0.195 & 0.373 & -0.276 & 0.296 & -0.323 & 0.226 & -0.370 \\
			\multicolumn{1}{c}{} & {$\beta_{1}$} & PVFCR & 0.485 & 0.894 & 0.253 & 0.788 & 0.192 & 0.765 & 0.136 & 0.737 \\
			\multicolumn{1}{c}{} &     & CR  & 0.499 & 0.914 & 0.270 & 0.814 & 0.211 & 0.793 & 0.155 & 0.768 \\
			\multicolumn{1}{c}{2} & {$\eta$} & PVFCR & 0.977 & 1.327 & 0.718 & 1.009 & 0.578 & 0.869 & 0.434 & 0.721 \\
			\multicolumn{1}{c}{} &     & CR  & 1.131 & 1.458 & 0.911 & 1.177 & 0.768 & 1.038 & 0.606 & 0.894 \\
			\multicolumn{1}{c}{} & {$\alpha$} & PVFCR & 0.377 & -0.222 & 0.268 & -0.195 & 0.224 & -0.166 & 0.191 & -0.149 \\
			\multicolumn{1}{c}{} &     & CR  & 0.557 & -0.479 & 0.443 & -0.399 & 0.391 & -0.355 & 0.350 & -0.326 \\
			\multicolumn{1}{c}{} & {$\lambda$} & PVFCR & 0.203 & 1.131 & 0.101 & 1.050 & 0.077 & 1.028 & 0.053 & 1.004 \\
			\multicolumn{1}{c}{} &     & CR  & 0.138 & 1.035 & 0.082 & 0.988 & 0.073 & 0.975 & 0.065 & 0.960 \\
			\multicolumn{1}{c}{} & {$\gamma$} & PVFCR & 0.393 & 0.511 & 0.327 & 0.579 & 0.299 & 0.610 & 0.265 & 0.647 \\
			\multicolumn{1}{c}{} & {$\sigma^2$} & PVFCR & 1.107 & 0.905 & 1.201 & 0.807 & 1.221 & 0.788 & 1.231 & 0.782 \\
			\hline
			\label{gama_09}
	\end{tabular}
 \end{scriptsize}
\end{table}%

\clearpage

\section{Application}
\label{aplicacao}

In this section, the proposed model and CR model are fitted to a real data set. The data are part of a study about cutaneous melanoma for the evaluating of postoperative treatment performance with a high dose of interferon alfa-2b drug in order to prevent recurrence. Patients were included in the study from $1991$ to $1995$ and follow-up was conducted until $1998$. The data were collected by \cite{Ibrahim} where survival time is defined as the time until the patient's death. The sample size is $m=417$ patients and the percentage of censored observations is $56\%$. The explanatory variables measured at baseline are: treatment (control or interferon), age (in years), sex, performance status (patient's functional capacity scale) and nodule category (categorization of number of lymph nodes:  category $1$ if $0$ lymph node, category $2$ if $1$ lymph node, category $3$ if $2$ or $3$ lymph nodes and category $4$ if lymph nodes $\geq 4$).

For fitted PVFCR and CR models, we considered $n_p=1000$, where the first $10000$ iterations were eliminated as burn-in samples and considered jump of size $100$.  The estimates of $\mbox{\boldmath{$\vartheta$}}$ and cure rate are obtained by (\ref{est_par_bayes}) and (\ref{est_p0_bayes}), respectively, and the estimator of improper survival function is given by (\ref{relia_bayes}).

\begin{figure}[!h]
	\centering
	\includegraphics[width=0.5\linewidth]{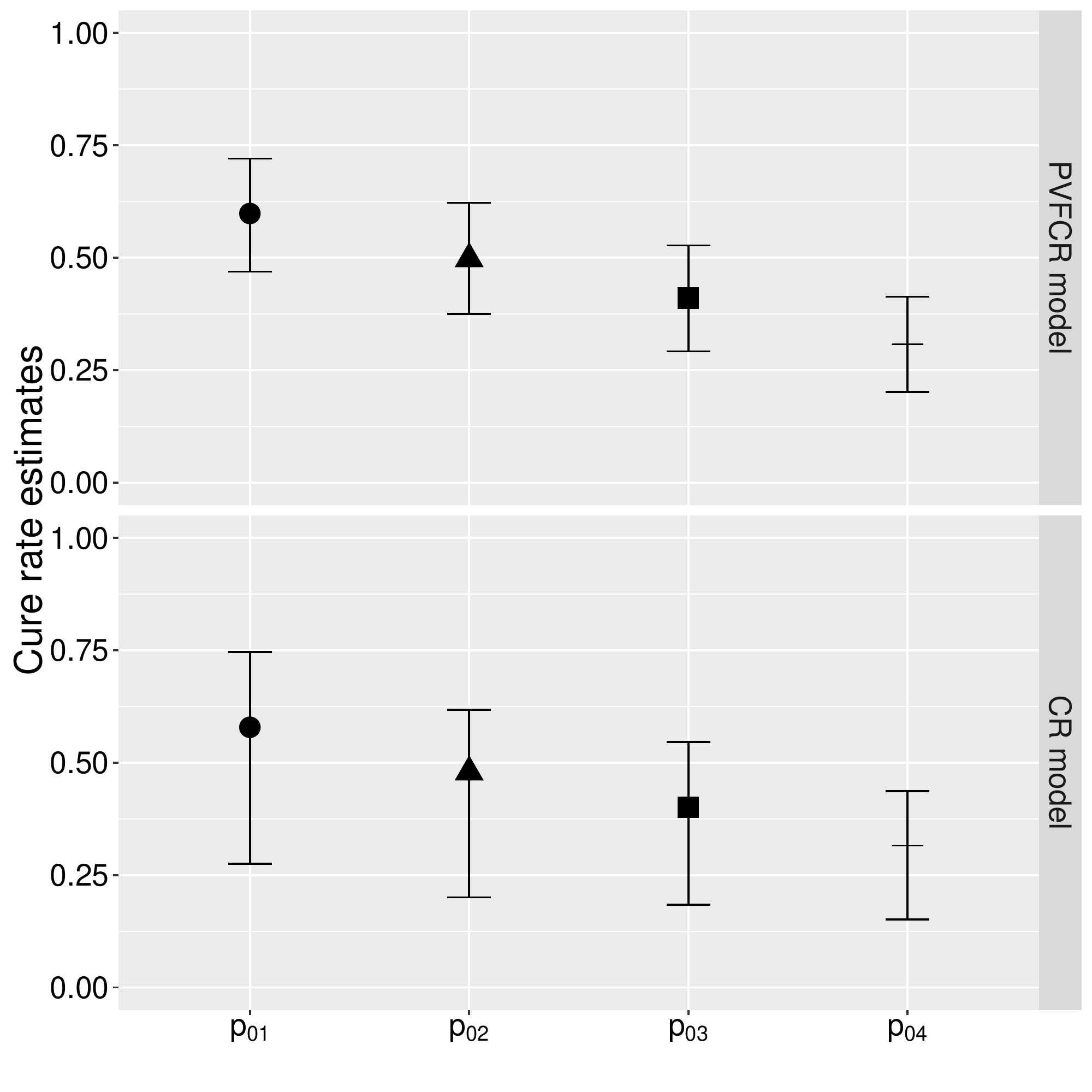}
	\caption{Cure rate estimates (symbol) and HPD interval (bar) according to fitted PVFCR and CR models.}
	\label{fig:prob_cura}
\end{figure}

\begin{figure}[!h]
	\centering
	\includegraphics[width=0.5\linewidth]{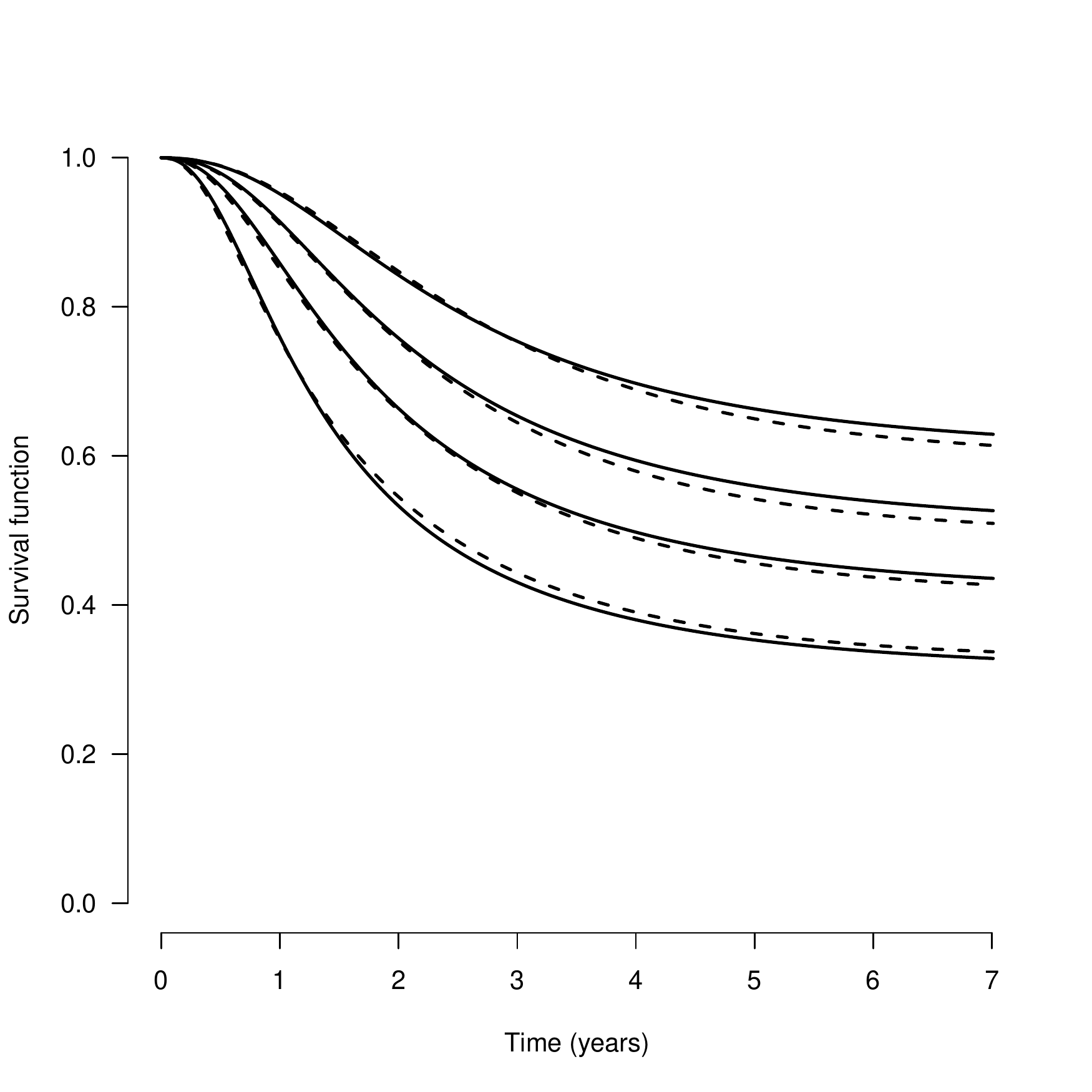}
	\caption{Survival estimates curves by PVFCR model (solid line) and CR model (dotted) stratified by nodule category (1-4 from top to bottom).}
	\label{fig:sobrevida}
\end{figure}

Except nodule category, all regression coefficients are non-significant for both fitted model. Then, in Table \ref{tabela_estimativas} is presented the summaries of parameters estimates of final model (considering only dummies variables of nodule category as explanatory variable, where the lowest category is baseline). We can note that the standard deviation of all the parameters are lower for the proposed model, as well as the HPD intervals have lower amplitudes. Furthermore, PVFCR model showed a slightly higher CPO value ($\mbox{CPO}=-516.4$ for PVFCR model versus $\mbox{CPO}=-516.6$ for CR model). Although the inference is the same for both models: only the explanatory variable nodule category is significant, the models provides similar fit for survival curves (Figure \ref{fig:sobrevida}) and category 1 is statistically different from categories $3$ and $4$ that have the lowest cure rate; HPD intervals of cure rates have lower amplitudes for PVFCR model, as we can observed in Figure \ref{fig:prob_cura}. Besides we emphasize the importance of the proposed model in capture and in quantifying the degree of unobservable heterogeneity.
\begin{table}[!h]
	\centering
	\caption{Parameters posterior mean, standard deviation (SD) and Highest Probability Density interval (HPD) of fitted PVFCR and CR models.}
	{\begin{tabular}{ccccccccccc} 
			\hline
			& \multicolumn{4}{c}{PVFCR model}                                           & & \multicolumn{4}{c}{CR model} \\ \cline{2-5} \cline{7-10}
			\raisebox{-1.7ex}[0pt][0pt] {Parameter}       & \raisebox{-1.7ex}[0pt][0pt] {Mean}    &   \raisebox{-1.7ex}[0pt][0pt] {SD}   & \multicolumn{2}{c}{HPD 95\%} & & \raisebox{-1.7ex}[0pt][0pt] {Mean}    &   \raisebox{-1.7ex}[0pt][0pt] {SD}    & \multicolumn{2}{c}{HPD 95\%}   \\
			&         &                         & Lower      & Upper               & &          &                         & Lower   &  Upper            \\ \hline
			$\lambda$	&	2.355	&	0.267	&	1.852	&	2.863	& & 	2.307	&	0.296	&	1.716	&	2.822	 \\
			$\alpha$	&	-3.147	&	0.856	&	-4.859	&	-1.802	& & 	-3.889	&	2.216	&	-8.070	&	-1.705	 \\
			$\eta$	&	2.919	&	1.330	&	0.596	&	5.651	& & 	3.670	&	1.461	&	0.939	&	6.414	\\
			$\beta_0$	&	0.233	&	0.743	&	-0.928	&	1.571	& & 	0.886	&	2.145	&	-0.901	&	4.816	 \\
			$\beta_{D2}^{\rm }$	&	0.674	&	0.396	&	-0.026	&	1.501	& & 	0.809	&	0.431	&	-0.039	&	 1.599	 \\
			$\beta_{D3}$	&	1.313	&	0.522	&	0.382	&	2.368	& & 	1.503	&	0.576	&	0.496	&	2.556	 \\
			$\beta_{D4}$	&	2.108	&	0.519	&	1.156	&	3.144	& & 	2.295	&	0.532	&	1.305	&	3.322	 \\
			$\gamma$	&	0.413	&	0.264	&	0.002	&	0.904	& & 	-	&	-	&	-	&	-	\\
			$\sigma^2$	&	1.270	&	1.067	&	0.007	&	3.378	& & 	-	&	-	&	-	&	-	\\
			\hline
			\end{tabular}}
		\label{tabela_estimativas}
		\\ \footnotemark{$\beta_{Dl}$ is the parameter associated to $l$th dummy variable that is indicates $l$th nodule category, for $l=2,3,4$ (category $1$ is baseline).}
\end{table}

\newpage

\section{Final remarks}
\label{conclusoes}

In this paper, we look at the cure rate model formulated by \cite{cancho2011flexible} in a different way, that is, we considered a random unobservable effect in promotion time of each competing cause, which allows to quantify the unobserved heterogeneity. The PVF frailty model was considered for the latent variables and it includes many other frailty models as special cases, being of great interesting. A simulation study was conducted to illustrate the good performance of the Bayesian estimators of the proposed model, where the RMSE appears reasonably close to zero as sample size increases. The results indicated lower RMSE for the estimators of the proposed model parameters, mainly in presence of large unobservable heterogeneity. As in practice situation the choice of the model is often based on a selection criterion, we evaluated the performance of model in terms of CPO criterion (higher values are desirable) when it is compared to usual cure rate model \cite{cancho2011flexible}. We observed that, in average, the CPO of fitted proposed model is largest, exception when $\gamma$ close to one. The practical relevance and applicability of the proposed model is demonstrated in a real data set, which our model yields a slight better fit than the usual cure rate model. We hope this generalization may attract wider applications in survival analysis. The computational codes can be requested for the first author.

\bibliographystyle{abbrvnat-m1}

\end{document}